\documentclass[useAMS,usenatbib]{mn2e}

\usepackage{graphicx}

\title[First measurements of the magnetic field on FK Com]{First measurement 
of the magnetic field on FK Com and its relation to the contemporaneous 
starspot locations\thanks{Based on observations collected at the European 
Southern Observatory, Chile (Prg. 280.D-5075); at the automatic STELLA 
observatory at Tenerife, Spain; and with the Vienna automatic photometric 
telescopes Wolfgang and Amadeus, Arizona, USA}}
\author[H. Korhonen et al.]{H. Korhonen$^1$\thanks{E-mail:hkorhone@eso.org},
S. Hubrig$^2$, S.V. Berdyugina$^{3,4}$, Th. Granzer$^5$, T. Hackman$^6$, 
\newauthor
M. Sch{\"o}ller$^1$, K.G. Strassmeier$^5$ and M.  Weber$^5$\\
$^1$European Southern Observatory, Karl-Schwarzschild-Str.\ 2, D-85748 
  Garching bei M\"unchen, Germany \\ 
$^2$European Southern Observatory, Casilla 19001, Santiago, Chile\\
$^3$Kiepenheuer Institut f\"ur Sonnenphysik, D-79104 Freiburg, Germany\\
$^4$Institute of Astronomy, ETH Z\"urich, 8093 Z\"urich, Switzerland\\
$^5$Astrophysikalisches Institut Potsdam, An der Sternwarte 16, D-14882 
     Potsdam, Germany\\
$^6$Observatory, PO Box 14, FI-00014 University of Helsinki, Finland}

\begin{document}

\date{Accepted 1988 December 15.
      Received 1988 December 14;
      in original form 1988 October 11}

\pagerange{\pageref{firstpage}--\pageref{lastpage}} \pubyear{2008}

\maketitle

\label{firstpage}

\begin{abstract}
In this study we present simultaneous low-resolution longitudinal magnetic 
field measurements and high-resolution spectroscopic observations of the cool 
single giant FK~Com. The variation of the magnetic field over the rotational 
period of 2.4\,days is compared with the starspot location obtained using 
Doppler imaging techniques, V-band photometry and V-I colours. The 
chromospheric activity is studied simultaneously with the photospheric activity
using high resolution observations of the H$\alpha$, H$\beta$ and H$\gamma$ 
line profiles. Both the maximum (272$\pm$24\,G) and minimum (60$\pm$17\,G) in 
the mean longitudinal magnetic field, $\left<B_z\right>$, are detected close 
to the phases where cool spots appear on the stellar surface. A possible 
explanation for such a behaviour is that the active regions at the two 
longitudes separated by 0.2 in phase have opposite polarities.
\end{abstract}


\section{Introduction}

The strongest magnetic fields in non-degenerate stars are measured in Ap stars 
(see, e.g., \citealt{bab, land, hub1}). The development of spectropolarimetric 
techniques in the last years has enabled us to detect the weaker stellar 
magnetic fields seen in many cool stars, and to study the topologies of their 
magnetic fields in detail using Zeeman-Doppler imaging.

Zeeman-Doppler imaging \citep{ZDI} is closely related to traditional Doppler 
imaging (see, e.g., \citealt{vogt_DI, pisk90}) where high resolution, high 
signal-to-noise (S/N) spectra at different rotational phases of a star are 
used to measure the rotationally modulated distortions in the line-profiles. 
These distortions are produced by the inhomogeneous distribution of the 
observed characteristic, e.g., effective temperature or elemental abundance. 
Surface maps, so called  Doppler images, are constructed by combining all the 
observations from different rotational phases and comparing them with synthetic
model line-profiles. In Zeeman-Doppler imaging similar techniques are used, 
but the observations consist of high resolution spectropolarimetric 
observations, either including circular polarisation (Stokes V) or preferably 
the full Stokes vector. Using this technique different types of stars have 
been mapped: young rapid rotators (e.g., \citealt{3stars, marsden_V889Her}), 
more evolved stars (e.g., \citealt{petit_HD199178, 3stars}), and M dwarfs 
(e.g., \citealt{Donati_V374Peg}).

Zeeman-Doppler imaging provides detailed maps of the stellar surface 
magnetic fields, but observationally this technique is very demanding as high 
resolution, high S/N observations in polarised light are needed. On the other 
hand, information on the longitudinal magnetic field can be obtained also 
using low resolution spectropolarimetric observations in circular 
polarisation (see, e.g., \citealt{hub_06}). With such observations we can 
retrieve the component of the magnetic field parallel to the line-of-sight, 
weighted by the local emergent spectral line intensity and averaged over the 
stellar hemisphere visible at the time of observation. Due to the low 
resolution, such spectropolarimetric observations, however, do not provide a 
detailed insight into the magnetic surface structure since the polarised 
line-profiles of temperature and magnetic field sensitive lines cannot be 
studied individually.   

Here we present the first study of the mean longitudinal magnetic field of 
FK~Com, which is the prototype of the small class of FK~Comae stars 
\citep{bopp_stencel}. These stars are magnetically very active late-type 
giants with photometric and spectroscopic characteristics similar to those of 
the very active RS~CVn stars. The main difference between the groups is that 
the RS~CVn stars are close binary systems in which the tidal effects produce 
synchronous rotation, and therefore also rapid rotation, whereas the FK~Comae 
stars do not show significant radial velocity variations caused by a 
companion, and are thus most likely single stars. 

FK~Com itself, with its $v$\,sin\,$i$ of 160\,km/s, is the fastest rotator of 
the FK~Comae stars, and hence also the most active among this group. The 
spectrum of FK~Com was first described by \citet{merrill}. He noted a large 
projected rotational velocity, H$\alpha$ and Ca II H\&K emission and the 
variability of the H$\alpha$ profile. Due to the very broad line profiles the 
radial velocity of FK~Com is difficult to measure very accurately, leaving 
still some room to speculate on its possible binarity. The radial velocity 
variations have been constrained to $\pm 3$\,km/s by \citet{hue93} giving very 
strict limits to the mass of the possible companion. Small visual brightness 
variations in FK~Com of $\sim$0.1\,mag in the V-band with a period of 2.412 
days were first reported by \citet{chu66}. These variations are interpreted to 
be caused by large starspots on the surface. Several surface temperature maps 
of FK~Com have been obtained over the years using Doppler imaging technique 
(see, e.g., \citealt{fkcom1, fkcom5}). These surface temperature maps mainly 
show high latitude spots with spot temperatures of typically 1000~K less than 
that of the unspotted surface.

The current work provides the first measurement of the magnetic field on 
FK~Com. The low resolution spectropolarimetric observations in circular 
polarisation are used to infer the behaviour of the mean longitudinal magnetic 
field at different rotational phases. The Doppler imaging technique is applied 
to the high resolution spectra to construct a simultaneous surface temperature 
map. The longitudinal magnetic field measurements are compared to the 
contemporaneous starspot locations obtained from the surface image and the 
chromospheric activity seen in the H$\alpha$, H$\beta$ and H$\gamma$ lines.

\section{Observations and data analysis}

Low-resolution spectropolarimetric observations were obtained at the European 
Southern Observatory with FORS\,1 (FOcal Reducer low dispersion Spectrograph, 
see \citealt{fors}) mounted on the 8-m Kueyen telescope of the VLT. For the 
Doppler imaging the high resolution spectra in integral light were observed at 
the fully robotic 1.2-m STELLA observatory in Tenerife (see, e.g., 
\citealt{kgs_stella, michi_stella}). Photometric observations in the V and I 
bands were obtained with the Vienna automatic photometric telescope Amadeus in 
Arizona (e.g., \citealt{kgs_APT, granz_APT}). The phases for all the 
observations were calculated using the ephemeris obtained from 25 years of 
photometric observations, 
HJD= 2\,439\,252.895 + ($2\fd4002466\pm0\fd0000056$)E, 
referring to a photometric minimum and period calculated by \citet{jetsu93}.

All the observations from different sites used in this work are overlapping in 
time, but they have rather different time spans. The high resolution 
spectroscopy from the STELLA observatory is taken over just two stellar 
rotations, whereas the spectropolarimetry with FORS\,1 is from almost six 
rotations and the photometric observations extend over 16 stellar rotations. 
The basis for using observations from so vastly varying time periods is that 
the spot configuration does not change significantly during this time. The 
stability of the surface structures is discussed in detail in 
Section~\ref{evol}.

\subsection{Spectropolarimetry}

FORS\,1 is a multi-mode instrument equipped with polarisation analysing optics
comprising of super-achromatic half-wave and quarter-wave phase retarder 
plates, and a Wollaston prism with a beam divergence of 22$\arcsec$ in standard
resolution mode. Spectropolarimetric observations were obtained between April 
12 and April 26, 2008 using  GRISM 600B and a 0.4$\arcsec$ slit, resulting in 
a resolving power ($\lambda/\Delta\lambda$) of $\sim$2000 and a wavelength 
coverage of 3250--6215\,\AA{}. We used a non-standard readout mode (200kHz, 
low, 1$\times$1), which provided a broader dynamic range, hence allowing us to 
increase the S/N of individual spectra. To minimise the cross-talk effect we 
executed the sequence of sub-exposures $+45-45$, $+45-45$, $+45-45$ etc., up 
to six times, and calculated the values $V/I$ using:
\begin{equation}
\frac{V}{I} =
\frac{1}{2} \left\{ \left( \frac{f^{\rm o} - f^{\rm e}}{f^{\rm o} + f^{\rm e}} \right)_{\alpha=-45^{\circ}}
- \left( \frac{f^{\rm o} - f^{\rm e}}{f^{\rm o} + f^{\rm e}} \right)_{\alpha=+45^{\circ}} \right\},
\label{eqn:one}  
\end{equation}
where $\alpha$ gives the position angle of the retarder waveplate and 
$f^{\rm o}$ and $f^{\rm e}$ are ordinary and extraordinary beams, 
respectively. Stokes $I$ values have been obtained from the sum of the 
ordinary and extraordinary beams. The typical exposure time for each 
sub-exposure was about 120~sec. The achieved S/N for the whole sequences of 
sub-exposures were in the range from 2200 to 2500. More details on the 
observing technique with FORS\,1 can be found, e.g., in a paper by 
\citet{Hubrig2004}. 

The mean longitudinal magnetic field is diagnosed from the slope of a linear 
regression of $V/I$ versus the quantity
\begin{equation}
-\frac{g_{\rm eff}e}{4\pi{}m_ec^2} \lambda^2 \frac{1}{I} \frac{{\mathrm d}I}{{\mathrm d}\lambda} \left<B_z\right> + V_0/I_0
\end{equation}
where $V$ is the Stokes parameter which measures the circular polarization,
$I$ is the intensity observed in unpolarised light, $g_{\rm eff}$ is the 
effective Land\'e factor, $e$ is the electron charge, $m_e$ the electron mass, 
$c$ the speed of light, $\lambda$ is the wavelength, 
${{\rm d}I/{\rm d}\lambda}$ is the derivative of Stokes $I$, and 
$\left<B_z\right>$ is the mean longitudinal field. Our experience from a study 
of a large sample of magnetic and non-magnetic Ap and Bp stars revealed that 
this regression technique is very robust and that detections with 
$B_z > 3\,\sigma$ result only for stars possessing magnetic fields (see, e.g., 
\citealt{hub_06b}). 

\begin{figure}
  \includegraphics[width=8cm]{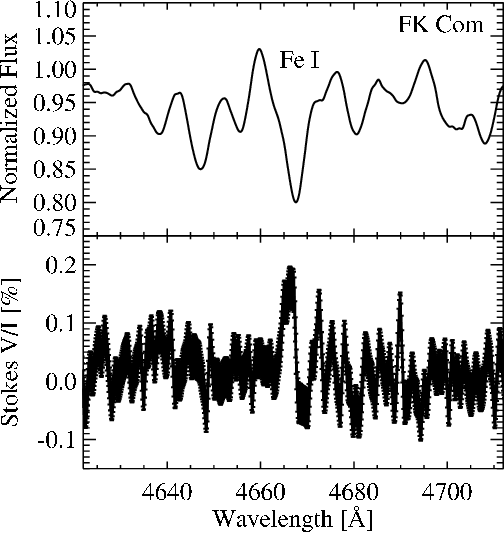}
  \caption{An example of FORS\,1 Stokes~I and V spectra of FK~Com in the 
    spectral region around the line Fe\,{\sc i} $\lambda$ 4668\,\AA{} at the 
    time of the maximum of the longitudinal magnetic field. The thickness 
    of the plotted lines in the Stokes~V spectra corresponds to the 
    uncertainty of the measurement of polarisation determined from photon 
    noise. 
}
  \label{stokes_spectra}
\end{figure}

\begin{figure}
  \includegraphics[width=8cm]{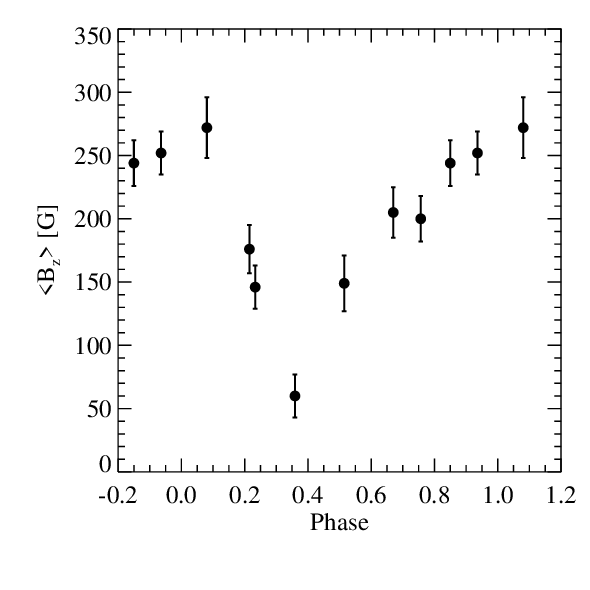}
  \caption{The mean longitudinal magnetic field $\left<B_z\right>$ plotted 
    over the rotational phase.}
  \label{mg_phase}
\end{figure}

In Fig.~\ref{stokes_spectra} we show an example of Stokes~I and V spectra of 
FK~Com in the spectral region around the line Fe\,{\sc i} 
$\lambda$ 4668\,\AA{} at the time of the field maximum of the longitudinal 
magnetic field. The values for the longitudinal magnetic field are measured 
using all the lines in the spectra. When $\langle$$B_z$$\rangle$ is plotted 
over the rotational phase a clear variation of the field is seen, with the 
maximum value $\langle$$B_z$$\rangle$\,=\,272$\pm$24\,G at the phase 0.08 and 
the field minimum  $\langle$$B_z$$\rangle$\,=\,60$\pm$17\,G at the phase 0.36 
(see Fig.~\ref{mg_phase}). In Table~\ref{fors_obs} we list the dates of 
observations, corresponding rotational phases and our measurements with 
1$\sigma$ uncertainties.

\begin{table}
  \caption[]{The mean longitudinal magnetic field measurements with 
    FORS\,1 over the rotational period. The quoted errors are 1$\sigma$ 
    uncertainties.} 
  \label{fors_obs}
  \centering
  \begin{tabular}{@{}cccr@{$\pm$}l@{}} \hline
    \multicolumn{1}{c}{Date} &
    \multicolumn{1}{c}{HJD} &
    \multicolumn{1}{c}{Phase} &
    \multicolumn{2}{c}{$\left<B_z\right>$} \\
    \hline
    12.04.08 & 54569.21944 & 0.938 & 252 & 17 \\
    13.04.08 & 54570.23517 & 0.361 & 60  & 17 \\
    15.04.08 & 54572.28969 & 0.217 & 176 & 19\\
    19.04.08 & 54576.21363 & 0.852 & 244 & 18 \\
    20.04.08 & 54577.13387 & 0.235 & 146 & 17 \\
    21.04.08 & 54578.18142 & 0.672 & 205 & 20 \\
    22.04.08 & 54579.16747 & 0.083 & 272 & 24 \\
    23.04.08 & 54580.20949 & 0.517 & 149 & 22 \\
    26.04.08 & 54583.19039 & 0.759 & 200 & 18 \\
    \hline
  \end{tabular}
\end{table}

\subsection{High resolution spectroscopy}

The high resolution spectroscopic observations used for Doppler imaging and 
investigating H$\alpha$ variations were obtained with the STELLA-I robotic 
telescope between April 14 and April 19, 2008. The fibre-fed SES spectrograph 
provides a resolving power ($\lambda/\Delta\lambda$) of 55\,000 and a 
wavelength coverage 3900--9000\,\AA{} in a fixed spectral format. The exposure 
time was always 3600 sec, and it provided an average S/N of 115 per resolution 
element in the Doppler imaging region around 6400\,\AA. The spectra were 
reduced using the SES reduction pipeline \citep{ses_red}. This pipeline is 
based on IRAF and it does the standard reduction steps, i.e., bias subtraction,
flat field correction, scattered light removal and optimal extraction of 
the spectral orders. Due to the overlapping orders, around 6000\,\AA{} and 
redder, the scattered light fit and the light pollution from adjacent orders 
is corrected using the HAMSCAT programme \citep{hamscat}.

A full list of observations together with the date, rotational phase, exposure 
time and S/N are given in Table~\ref{STELLA_obs}.

\begin{table}
  \centering
  \caption[]{The high resolution spectroscopy with SES at the STELLA 
    observatory. The date of the observations, rotational phase and the S/N 
    per resolution element are given.}
  \label{STELLA_obs}
  \centering
  \begin{tabular}{@{}cccrc@{}}
    \hline
    Date & HJD & Phase & S/N \\ \hline
    14.04.08 & 2454571.37891 & 0.046 & 116 \\
    15.04.08 & 2454571.51563 & 0.103 & 100 \\
    15.04.08 & 2454571.63672 & 0.153 & 116 \\
    15.04.08 & 2454572.37891 & 0.462 & 124 \\
    16.04.08 & 2454572.66797 & 0.583 & 101 \\
    16.04.08 & 2454572.73047 & 0.609 & 99 \\
    16.04.08 & 2454573.37891 & 0.879 & 126 \\
    16.04.08 & 2454573.48828 & 0.925 & 121 \\
    17.04.08 & 2454573.65625 & 0.995 & 134 \\
    17.04.08 & 2454574.37891 & 0.296 & 126 \\
    18.04.08 & 2454574.58594 & 0.382 & 109 \\
    18.04.08 & 2454574.72656 & 0.441 & 84 \\
    18.04.08 & 2454575.40625 & 0.724 & 132 \\
    19.04.08 & 2454575.58594 & 0.799 & 119 \\
    19.04.08 & 2454575.69922 & 0.846 & 123 \\ \hline
  \end{tabular}
\end{table}

\begin{figure*}
  \includegraphics[width=18cm]{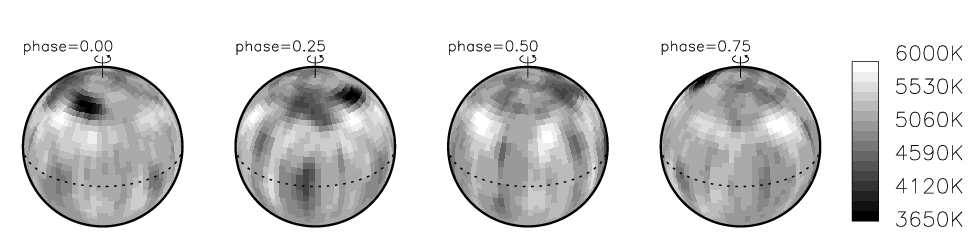}
  \caption{The surface temperature map of FK~Com. The surface is shown at 
	four different rotational phases which are 0.25 in phase apart. The 
	gray scale gives the temperature in Kelvin.}
  \label{DI}
\end{figure*}

\section{Surface temperature mapping}

A surface temperature map of FK~Com was obtained using Doppler imaging 
technique. For the imaging the Tikhonov Regularization code INVERS7PD, 
originally developed by N. Piskunov (see e.g., \citealt{pisk90}) and modified 
by T. Hackman \citep{hack_etal} was used. The stellar parameters adopted for 
Doppler imaging, and more details on the selection of the parameters, inversion
technique and line-profile calculation are given by \citet{fkcom1, fkcom5}. In 
the inversions the whole spectral region 6416--6444\,\AA{} was used, only 
the atmospheric line at 6433\,\AA{} was masked out. The temperature of the 
unspotted surface was set to 5200\,K. This is somewhat higher than what is 
usually derived for FK~Com, i.e., $5000\pm50$\,K. The difference could be 
caused by problems with scattered light removal or by noisiness of the data. 


The surface temperature map based on the STELLA data is shown in 
Fig.~\ref{DI}. The map shows two main active regions on the surface. One 
active region is spanning the phases 0.00--0.14 at an average latitude of 
58$^{\circ}$. This region has a minimum temperature of 3650~K, and it harbours 
the coolest spot seen on the surface for this time period. The other major spot
concentration is located at phases 0.21--0.35. This active region has a 
minimum temperature of approximately 4300~K and it spans two latitude ranges, 
one at high latitudes with an average latitude of 61$^{\circ}$ and another one 
at the equator.

The case where two high latitude spots of different contrast are approximately 
0.2 in phase apart, is commonly seen on the surface of FK~Com (see e.g., 
\citealt{fkcom5}). The large temperature difference of approximately 1500\,K 
between the coolest spot and the unspotted surface in the present map is 
larger than the average difference seen on FK~Com. Still, similar temperature 
differences have been reported before for some time periods, e.g., August 2002 
\citep{fkcom5}. On the whole, the spot configuration seen in the current map 
is similar to what one would expect on FK~Com based on the earlier Doppler 
imaging results. 

\section{Discussion}

\subsection{Spot evolution in April-May 2008}
\label{evol}

\begin{figure}
  \centering
  \includegraphics[width=8cm]{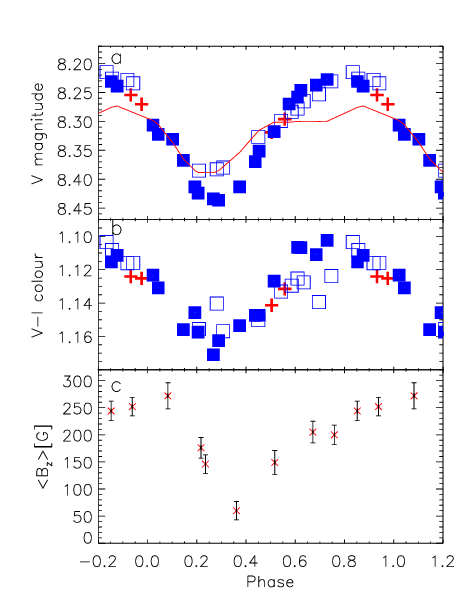}
  \caption{The V band magnitudes, V-I colour and mean longitudinal magnetic 
    field strength plotted against the rotational phase. a) V-band light-curve.
    The crosses denote observations obtained during the same time period as 
    the magnetic field measurement, filled squares show the observations
    obtained 11 days prior the magnetic field observations, and the open 
    squares display observations from the following 13 days. The solid line 
    gives the photometry calculated from the Doppler image, and shifted to the 
    average magnitude of the V band observations obtained before the 
    $\left<B_z\right>$ measurements. b) V-I colour. The symbols are the same 
    as for the V-band light-curve. c) The mean longitudinal magnetic field.}
  \label{MG_DI}
\end{figure}

The V-band light-curve is shown in Fig.~\ref{MG_DI}a. It has a broad minimum 
spanning phases 0.1--0.4. In the V light-curve the crosses are from the same 
time period as the $\left<B_z\right>$ measurements, filled squares from the 11 
days preceding the magnetic field measurements, and the open squares from the 
13 days following the field measurements. Some spot evolution is seen in the 
light-curve within the 38 days that the observations cover. The minimum gets 
broader and more shallow with time. Unfortunately, we have only four 
measurements coinciding with the magnetic field measurements, thus it is 
impossible to determine exactly the shape of the photometric minimum which 
corresponds to our magnetic field determination. Still, even though the exact 
shape of the light-curve minimum has somewhat changed, the locations of the 
minimum and maximum have not changed over the more than a month that the 
photometric observations cover. Thus, also our spectropolarimetric and 
spectroscopic observations that are overlapping in time, but spanning 14 and 4 
days, respectively, can be studied together.

The solid line plotted in Fig.~\ref{MG_DI}a is the V band photometry calculated
from the Doppler image, and shifted to the average magnitude of the V band 
observations obtained before the magnetic field measurements. Clearly, 
the shape of the calculated light-curve follows quite well the observed one. 
The main difference is detected in the amplitude of the light-curve, i.e., the 
calculated curve has smaller amplitude. This indicates that the spots around 
0.00-0.35, based on the photometry, should be larger or cooler than in the 
Doppler image. Another possible explanation for the small amplitude in the 
calculated light-curve is that the areas around phases 0.5--0.8, which in the 
Doppler image are 400--500~K cooler than the unspotted surface, and cause the 
dip seen in the calculated light-curve at these phases, are actually 
artifacts. Overall, the comparison of the light-curve calculated from the 
Doppler image and the observed V-band light-curve indicates that the location 
of the main cool spots seen in the temperature map are correct.

The V-I colour curve, also obtained by Amadeus, is shown in Fig.~\ref{MG_DI}b. 
It displays a similar shape as seen in the $\left<B_z\right>$ variation curve 
and the V-band light-curve. Unlike the V-band, the V-I colour does not show 
large differences in the shape of the minimum between the observations 
obtained before and after the magnetic field determination. This indicates 
that the relative temperature of the spotted regions remains the same. On the 
other hand, the \mbox{V-I} colour obtained before and after the 
$\left<B_z\right>$ measurements show differences around the phases 0.6--0.8. 
From the V-band light-curve it can be seen that at these phases the star is 
consistently slightly fainter after our magnetic field observations than 
before them. This behaviour, together with the behaviour seen in the V-I 
colour, indicates that there could be a new active region forming around 
phases 0.6--0.8.

\subsection{Correlation between the magnetic field measurements and the 
starspot location}

\begin{figure}
  \centering
  \includegraphics[width=4cm]{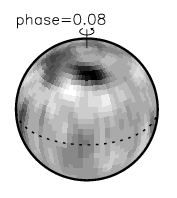}
  \includegraphics[width=4cm]{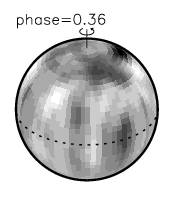}
  \caption{As Fig.~\ref{DI}, but now showing the surface of FK~Com at the 
    phases of the maximum and minimum $\left<B_z\right>$, phases 0.08 and 0.36 
    respectively}
  \label{DI_spots}
\end{figure}

As presented in Fig.~\ref{mg_phase}, the mean longitudinal magnetic field 
$\left<B_z\right>$ shows clear variations with the rotational phase. In 
Fig.~\ref{MG_DI}  the magnetic field strength is plotted together with the  
V-band light-curve and the V-I colour curve. The photometric minimum seen in 
the V-band light-curve, V-I colour, and also determined from the Doppler 
image, occurs close to the minimum strength of the mean longitudinal magnetic
field. The exact location of the $\left<B_z\right>$ minimum appears slightly 
shifted with respect to the photometric minimum, and also the maximum of 
$\left<B_z\right>$ occurs within the broad photometric minimum. 
Fig.~\ref{DI_spots} shows the surface temperature map centred at the two 
rotational phases that exhibit the $\left<B_z\right>$ maximum and minimum. 
These plots show clearly that the coolest spot occurs at the phase of the 
field maximum, i.e., at phase 0.08, and that the field minimum of phase 0.36 
coincides with the secondary active region seen on the surface.

It seems that the behaviour of $\left<B_z\right>$ in FK~Com is correlated with 
the starspots seen on the surface, and that the main spot on the surface, at 
phase 0.1, has positive magnetic field polarity. The minimum of the magnetic 
field strength occurs around the active region of phase 0.3. The observed 
minimum could be explained by this active region having a negative magnetic 
field polarity. The negative field would partly cancel the positive field 
emerging in the dominant spot. This cancellation becomes more efficient when 
the dominant active region moves closer to the stellar limb, and the secondary 
active region approaches the centre of the disk. Therefore, the minimum of 
$\left<B_z\right>$ is seen slightly shifted from the centre of the secondary 
active region, and also away from the dominant spot.

If the spot at the phase 0.1 has different polarity than the ones at 0.3, then 
this configuration closely resembles what is commonly observed in the sunspots.
In the Sun spots usually occur in pairs of different polarity, and the leading 
spot is compact and the following one fragmented. Thus, in Doppler images the 
leading spot would appear cooler, and the following one would have higher 
average temperature, as is also seen in the surface temperature map of FK~Com. 
In the Sun the spots are also connected by magnetic loops. Using Chandra X-ray 
observations \citet{drake} find some indication for magnetic loops also on 
FK~Com. Their observations suggest that the observed X-ray emission originates 
from plasma residing predominantly in extended structures centered at a phase 
halfway between two active regions, and that the coronal structures 
revealed by the Chandra observations correspond to magnetic fields joining 
these two active regions.

\subsection{Chromospheric activity}

\begin{figure}
  \centering
  \includegraphics[width=8.3cm]{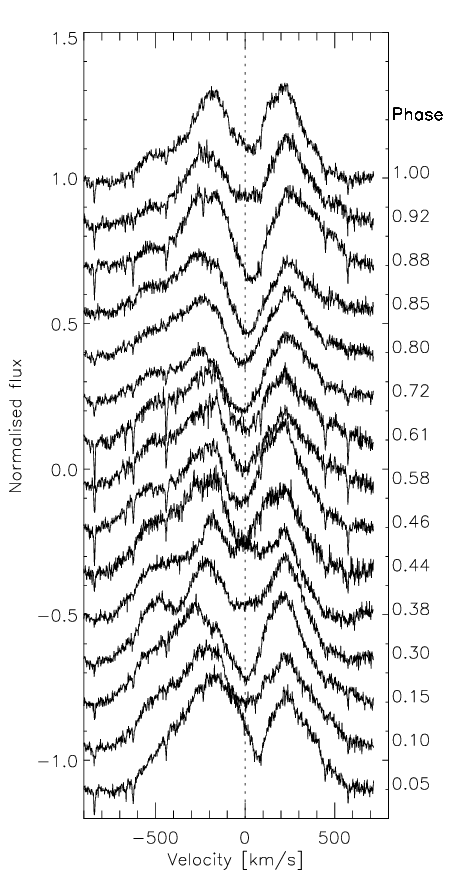}
  \caption{The H$\alpha$ line variations in FK~Com. All the 15 individual
    H$\alpha$ observations obtained in this study are shown plotted against 
    the velocity obtained in relation to the rest wavelength of the
    H$\alpha$.}
  \label{Ha}
\end{figure}

For studying the possible connection between the chromospheric and 
photospheric activity on FK~Com in April 2008, the H$\alpha$ line profiles 
obtained with STELLA~1 were also investigated. All the profiles from the 15 
epochs observed in this study are plotted in Fig.~\ref{Ha}. The radial velocity
of FK~Com, \mbox{-24~km/s} \citep{hue93}, has been removed from the profiles. 
In FK~Com the H$\alpha$ line is a very strong and wide double peaked emission 
line. The shape of the profile varies strongly with time, and especially the 
strength of the absorption core shows significant changes. 

Earlier studies have explained the observed behaviour of H$\alpha$ by 
structures similar to solar prominences. Based on the velocity behaviour of 
the Balmer lines studied by \citet{hue93}, the material in the prominences is 
confined within one stellar radius from the rotation axis. \citet{welty_etal} 
agree on the prominence theory and conclude also that their measurements can 
be explained by one single dominant emitting prominence. The work by 
\citet{oliv} also supports the existence of prominences, but they find 
evidence for several different prominences existing simultaneously. On the 
other hand, \citet{kjurk} suggest that the origin of the H$\alpha$ emission in 
FK~Com is a disk, which is half-illuminated. They also suggest that the source 
of illumination is a low-mass hot secondary orbiting the disk.

\begin{figure}
  \centering
  \includegraphics[width=8.3cm]{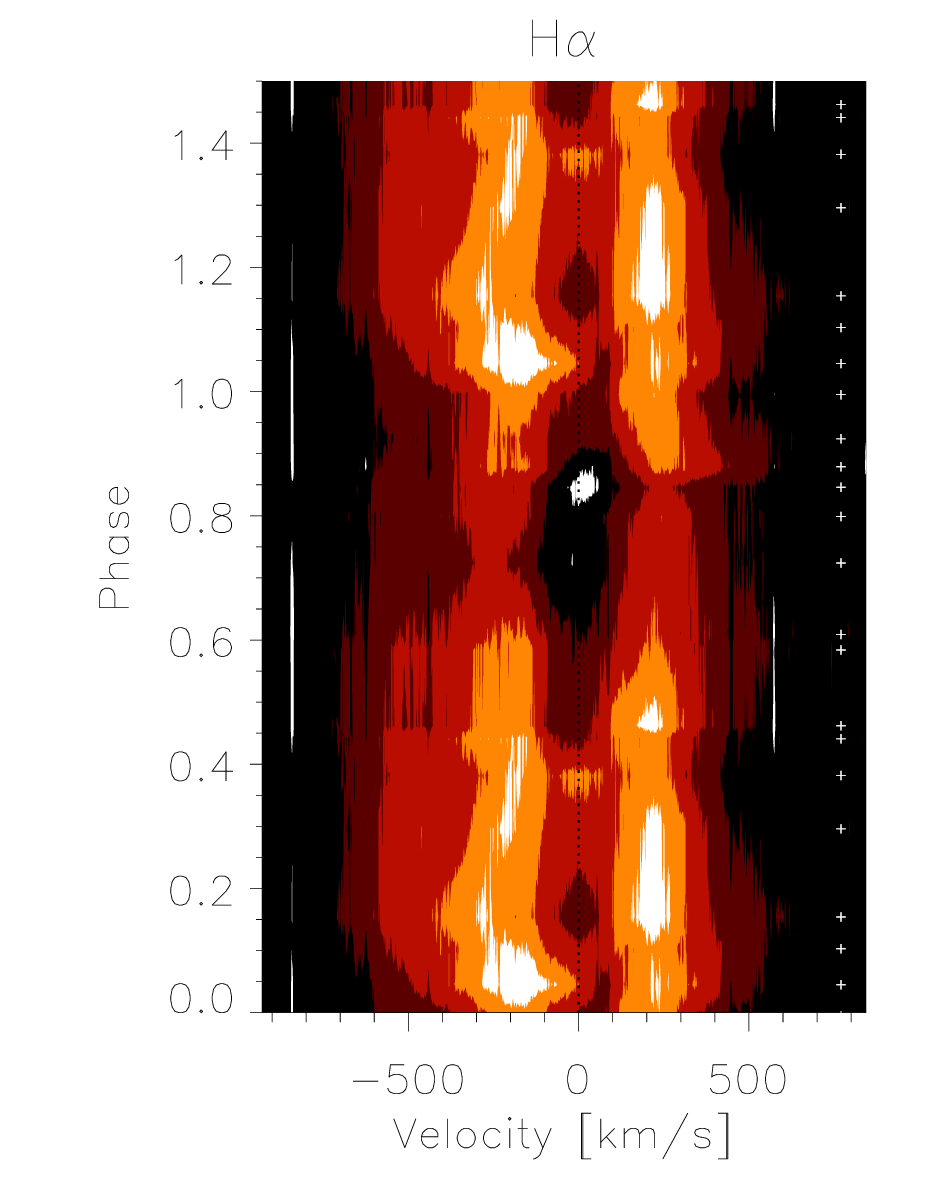}
  \caption{Dynamic spectrum of the H$\alpha$ line of FK~Com for April 2008. 
    The plot is shown from the phase 0.0 until 1.5 to better present the 
    structure around the phase 1.0. The crosses on the right hand side of the 
    plot give the phases of the observations and the dashed line the zero 
    velocity. Brighter the plot the more enhanced the emission is.}
  \label{Ha_dyn}
\end{figure}

\begin{figure}
  \centering
  \includegraphics[width=8.3cm]{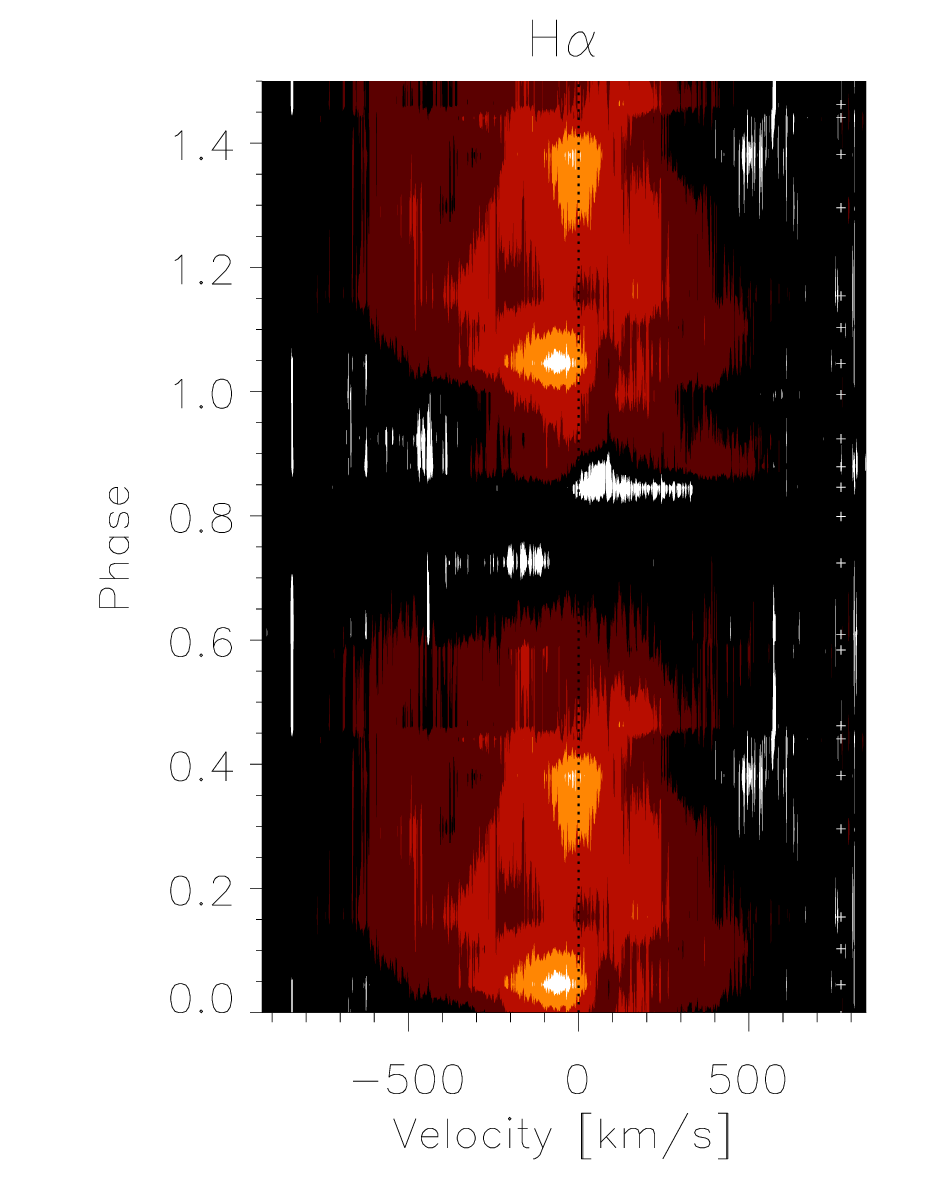}
  \caption{As in Fig.~\ref{Ha_dyn}, but the profile showing the minimum 
    chromospheric activity, phase 0.80, has been subtracted from all the 
    profiles before creating the dynamic spectrum.} 
  \label{Ha_dyn_min}
\end{figure}

For studying in detail the activity seen in the H$\alpha$ line a dynamic 
spectrum was constructed from the H$\alpha$ profiles. This spectrum is shown 
in Fig.~\ref{Ha_dyn}. Brighter the plot the more enhanced the emission is. The 
phases of the observations are shown with crosses on the plot. The data for 
the phases where there are no observations are interpolations between the 
closest phases with data. For showing the emission features better another 
dynamic spectrum (Fig.~\ref{Ha_dyn_min}) was created in which the profile 
showing the least chromospheric activity, phase 0.80, was subtracted from all 
the profiles.

The dynamic spectra show clearly that the phases close to the photospheric 
spots, i.e., 0.0--0.35, exhibit enhanced emission. At the phase 0.38 almost the
whole absorption core gets filled with emission (clearly seen also in 
Fig.~\ref{Ha}). This could be caused by a flare event originating close to the 
spot group around the phases 0.21--0.35, or by plage region rotating to the 
centre of the disk. As the HeI D$_{3}$ line (5876\AA), which is often seen in 
emission in stellar flares (see, e.g., \citealt{hue_ram}; \citealt{lopez}), 
does not show any emission during the observations, the flare is less likely 
option. Additionally, the H$\alpha$ profile at the phase 0.05 shows 
very strong emission, especially in the blue emission peak which extends to 
the red of the rest wavelength of H$\alpha$. Another intriguing structure in 
the line profiles is the increased emission at the velocity -550~km/s. This 
feature is very well seen at the phases 0.3--0.4, and is also present, to some 
extent, at most phases. It is also worth noting that the blue emission peak is 
wider than the red one, and often extends to velocities of -700 km/s, whereas 
the red peak usually only extends up to +500~km/s. The phases 0.60--0.85, 
during which the photospheric spots are hardly seen, show much less H$\alpha$ 
emission than the other phases, and at these phases the central absorption core
is also at its strongest. 

\begin{figure}
  \centering
  \includegraphics[width=4.15cm]{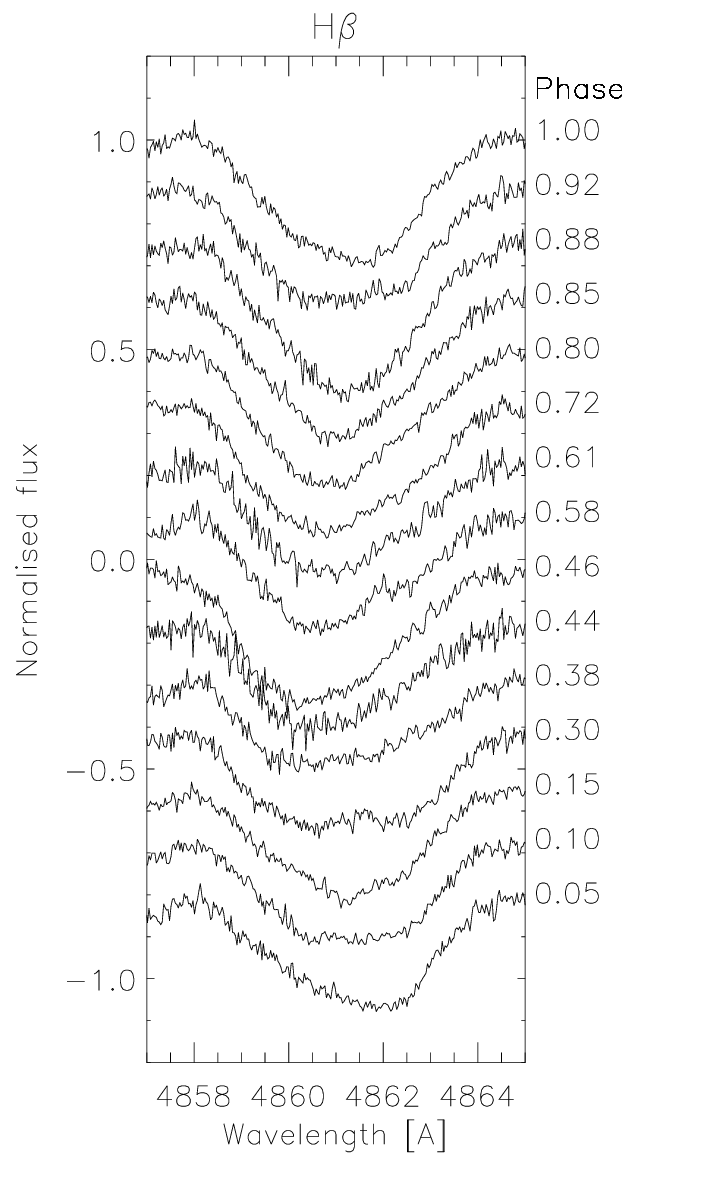}
  \includegraphics[width=4.15cm]{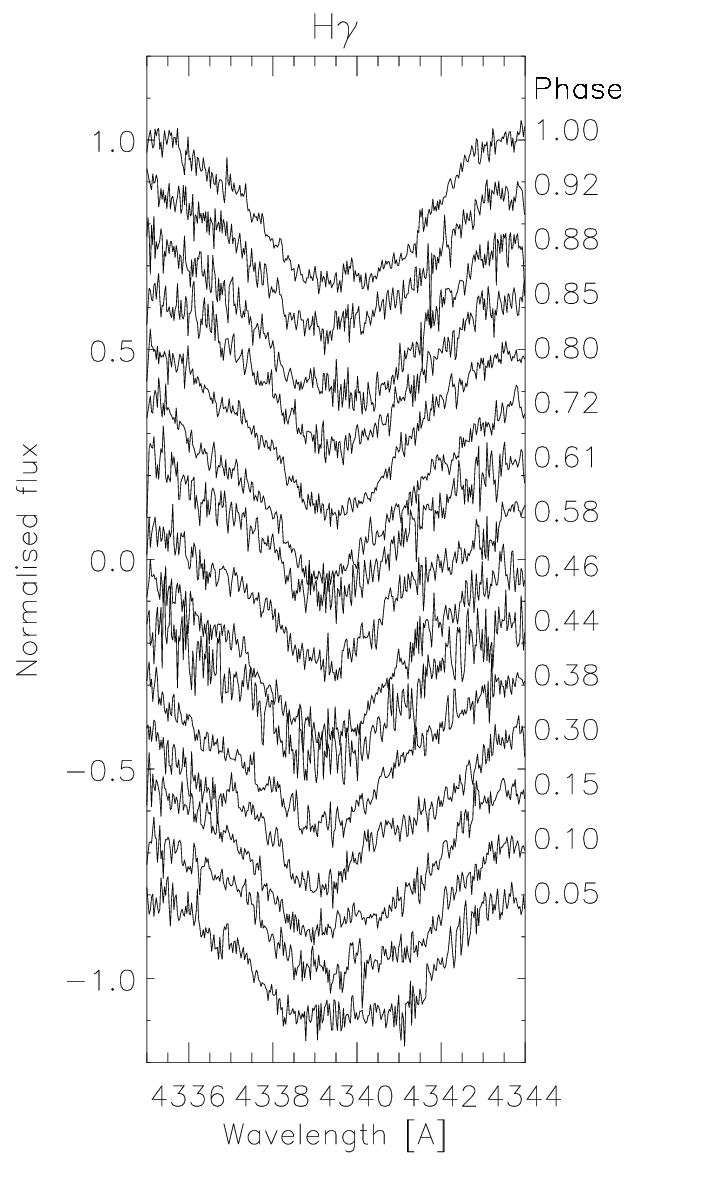}
  \caption{The H$\beta$ and H$\gamma$ line variations in FK~Com at all the 15 
    phases.} 
  \label{Hb_prof}
\end{figure}

\begin{figure}
  \centering
  \includegraphics[width=4.15cm]{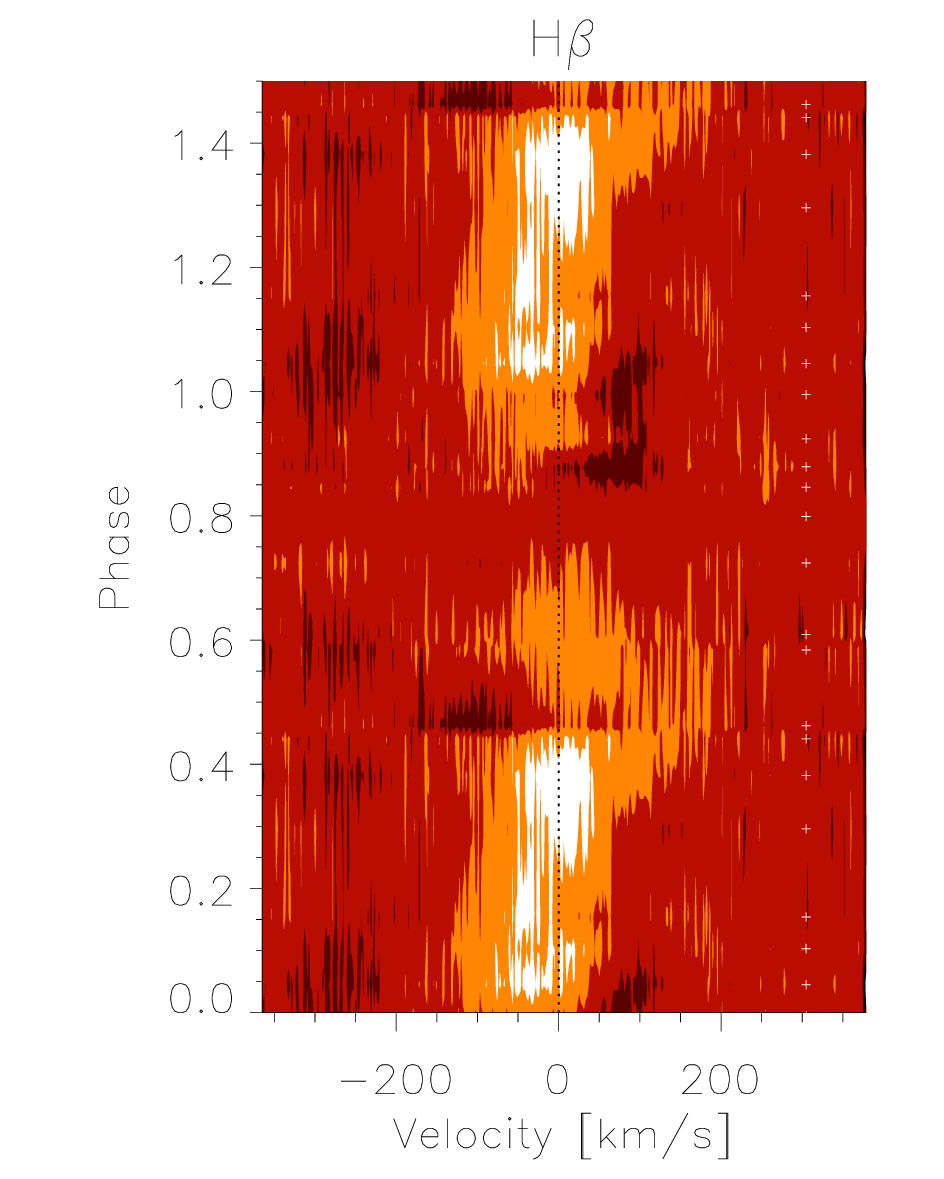}
  \includegraphics[width=4.15cm]{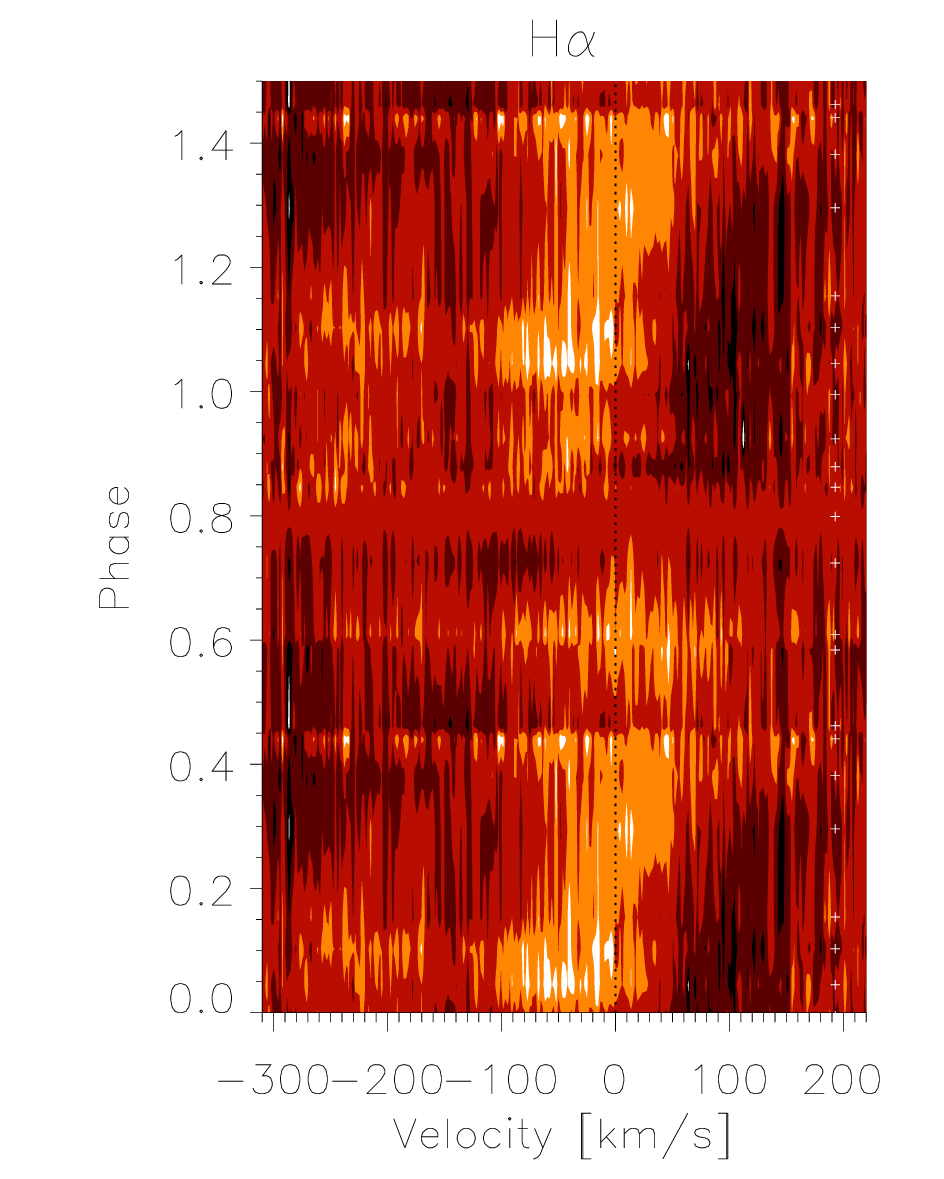}
  \caption{Dynamic spectrum of the excess emission in the H$\beta$ and 
    H$\gamma$ line. In both cases the profile showing the minimum chromospheric
    activity, phase 0.80, has been subtracted from all the profiles to create 
    the excess emission profiles.}
  \label{Hb_dyn}
\end{figure}

To further investigate the chromosphere of FK~Com, also the behaviour of 
H$\beta$ and H$\gamma$ lines was investigated. As can be seen from 
Fig.~\ref{Hb_prof}, in FK~Com H$\beta$ and H$\gamma$ are absorption lines that 
show temporal variations. In the case of H$\gamma$ the S/N is quite poor and 
the variations are not as easily seen as in H$\beta$. In both cases, the 
line is the deepest at phase 0.80, thus indicating the least amount of 
emission filling in the line core. When this line profile is subtracted from 
the other profiles, the excess emission at different phases can be studied. 
Fig.~\ref{Hb_dyn} shows the dynamical spectrum of the H$\beta$ and H$\gamma$ 
excess emission. Both pictures are strikingly similar to the one seen in the 
case of H$\alpha$, especially in H$\beta$ where the S/N is better. The phases 
where the photospheric spots are seen exhibit the most excess emission, and 
the chromosphere at the least spotted phases is again the least active.

\begin{figure*}
  \centering
  \includegraphics[width=10cm]{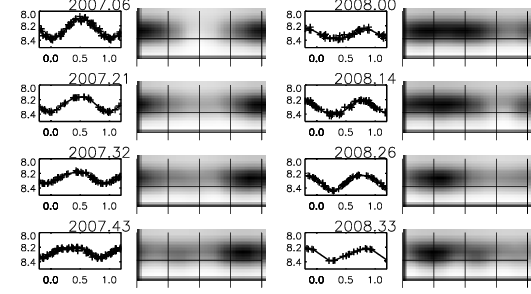}
  \caption{The light-curve inversion results obtained from the 2007--2008
    V-band photometry. The left panels show the observed light-curve
    (crosses) and the inversion result (solid line). The right panels are the
    spot filling-factor maps, where a darker colour indicates larger filling
    factor. The equator and the phases 0.0, 0.25, 0.50, 0.75 and 1.0 are
    shown by solid lines in the filling factor maps.}
  \label{fotoinv}
\end{figure*}

The H$\alpha$, H$\beta$ and H$\gamma$ observations show clearly that the 
chromosphere of FK~Com is more active at the phases where the photospheric 
spots are located. In H$\alpha$ the strong double peaked emission line is 
still present also at the phases in which the spots are not seen, but in these 
phases the emission peaks are less prominent and the absorption core is the 
deepest. On the whole, the H$\alpha$ profile of the non-spotted phases has a 
shape that can be explained by a disk, and the spotted phases exhibit enhanced 
emission, seen also in the H$\beta$ and H$\gamma$ profiles, which could 
originate from plages and prominences.

\subsection{The magnetic field and the flip-flop phenomenon}

In many active stars the spots concentrate on two permanent active
longitudes, which are 0.5 in phase apart (see, e.g., \citealt{ber_tuo}). In 
some of these stars the dominant part of the spot activity concentrates on one 
of the active longitudes, and abruptly switches the longitude every few years. 
This so-called flip-flop phenomenon was first discovered in FK~Com 
\citep{jetsu93}. Since the discovery, flip-flops have been reported in  
different types of stars, e.g., young solar type stars (e.g., 
\citealt{jar_ABDor}), RS~CVn binaries (e.g., \citealt{ber_tuo}) and the Sun 
\citep{ber_uso, uso}, even though this last result is still being debated 
\citep{pelt1, pelt2}.

The V-band photometry from 2007--2008 indicates that the latest 
flip-flop on FK~Com has happened during the latter part of the year 2007. In 
Fig.~\ref{fotoinv} the V-band light-curves are shown together with the results 
from the light-curve inversions (see \citealt{fkcom4}). The results for 2007 
show the main spot around phase 0.0 (or 1.0). In the last light curve inversion
from 2007, time period 2007.43, an indication of a spot forming approximately 
0.5 in phase apart from the main spot is seen. The first light-curve of 2008, 
2008.00, which was obtained during December 2007 and January 2008 shows a 
broad minimum around phase 0.0-0.5. During the beginning of 2008 the minimum 
became more concentrated and in the data from the time period 2008.26 one 
concentrated spot at the phase 0.25 is seen. This indicates that the active 
region observed between July 2007 and April 2008 has changed phase by 0.5 
(from phase 0.75 to 0.25), and thus forming a flip-flop event.

The flip-flop phenomenon has strong implications for the dynamo theory. The
behaviour of the global solar magnetic field can be explained by an 
axisymmetric mean-field dynamo without any longitudinal structure. In  
rapidly rotating stars the higher order non-axisymmetric modes become excited, 
forming two active regions that are 0.5 in phase apart (e.g., 
\citealt{moss95}), i.e., equivalent to the two permanent active longitudes. 
Not only is the magnetic configuration different in the axisymmetric and 
non-axisymmetric modes, but so are their oscillatory properties. The solar 
type axisymmetric modes show clear cyclic behaviour, whereas the 
non-axisymmetric modes do not oscillate. To explain the flip-flop phenomenon 
both properties, non-axisymmetric field configuration and oscillations, are 
needed. Recently, some attempts on modeling this behaviour have been published 
(e.g., \citealt{moss04, elst_kor}). From a theoretical point of view one of 
the key questions on the flip-flop phenomenon is: Do spots on the two 
permanent active longitudes have different magnetic field polarities?

Our current spectropolarimetric observations suggest the existence of two cool
spots seen on the surface having different polarities. Still, these two spots
have to be considered to be located on the same active longitude as they
together form the photometric minimum. Based on the photometric observations 
the second active longitude should be around phase 0.75, and at that phase 
no negative magnetic field polarity is seen. Although, a very small dip in 
$\left<B_z\right>$ is visible around the phase 0.75 which could be caused by a 
weak negative polarity field slightly canceling the signal from the positive 
polarity.

One has to keep in mind that the observations presented in this work 
have low spectral resolution and do not provide a detailed insight into the 
magnetic surface structure as polarised line-profiles of temperature and 
magnetic field sensitive lines cannot be studied individually. For properly 
answering the question on the spot polarities at the two permanent active 
longitudes, high resolution spectropolarimetric observations would be needed. 
Also, these high resolution observations should be carried out before and 
after a flip-flop event to study the possibility of the active longitudes 
changing polarity during the flip-flop, as suggested by \citet{tuo02}.

\section{Conclusions}

The following conclusions can be drawn from the low resolution 
spectropolarimetry, high resolution spectroscopy and broad band photometry 
presented in this work:

\begin{itemize}
\item The mean longitudinal magnetic field $\left<B_z\right>$, measured in 
April 2008, shows clear variations over the rotational phase of FK~Com, 
with a maximum $\langle$$B_z$$\rangle$\,=\,272$\pm$24\,G at the phase 0.08 and 
the minimum $\langle$$B_z$$\rangle$\,=\,60$\pm$17\,G at the phase 0.36.
\item The contemporaneous determination of the starspot locations by the 
Doppler imaging technique reveals three main spot groups on the surface of 
FK~Com in April 2008. Two groups are located at high latitudes close to 
60$^{\circ}$, and one is located close to the equator. The separation of the 
high latitude spot groups is 0.2 in phase.
\item Both the maximum and minimum value of $\left<B_z\right>$ occur within 
the phases of the broad photometric minimum. The maximum value coincides with 
the cool high latitude spot spanning the phases 0.00--0.14 and the minimum 
value with the high latitude and equatorial spots around the phases 
0.21--0.35. This behaviour could be explained by the active region around the 
phase 0.1 having positive magnetic field polarity and the less cool spots 
around phase 0.3 a magnetic field of negative polarity. 
\item The H$\alpha$ profiles show enhanced emission at the phases 0.0--0.4, 
coinciding with the photospheric spots. The phases 0.6--0.9, when the 
photospheric spots are hardly seen, show least emission and the strongest 
central absorption core. Similar behaviour is also seen in the H$\beta$ and 
H$\gamma$ lines. 
\item We would like to emphasise that obtaining high resolution 
spectropolarimetric observations of FK~Com, preferably before and after a 
flip-flop event, would provide valuable observational constraints for 
flip-flop Dynamos.
\end{itemize}

\section*{Acknowledgments}

We acknowledge ESO Director's Discretionary Time for allocating our observing 
run with FORS\,1 at the VLT. We thank the anonymous referee for his/her 
valuable comments that helped to improve the paper. We would also like to 
thank Katalin Ol{\'a}h for her careful reading and commenting of this paper.
SB acknowledges the EURYI Award from the European Science Foundation (see 
www.esf.org/euryi).

\label{lastpage}

\end{document}